\documentclass[sigconf]{acmart}
\usepackage{amsmath}
\usepackage{hyperref}
\usepackage{color}
\usepackage{algorithm}
\usepackage{algorithmic}
\usepackage{setspace}
\usepackage{natbib}
\usepackage{enumitem}
\usepackage{listings}
\usepackage{graphicx}
\usepackage{subcaption}
\usepackage{multirow}
\usepackage{booktabs}
\usepackage{tabularx}
\usepackage{stfloats}
\usepackage{tikz}
\usepackage{tcolorbox}
\usepackage{mdframed}
\usepackage{xcolor,colortbl}

\definecolor{c1}{cmyk}{0,0.6175,0.8848,0.1490}
\definecolor{c2}{cmyk}{0.1127,0.6690,0,0.4431}
\definecolor{c3}{cmyk}{0.3081,0,0.7209,0.3255}
\definecolor{c4}{cmyk}{0.6765,0.2017,0,0.0667}
\definecolor{c5}{cmyk}{0,0.8765,0.7099,0.3647}

\definecolor{lightgrey}{rgb}{0.93,0.93,0.93}

\newtcbox{\hlprimarytab}{on line, rounded corners, box align=base, colback=c3!10,colframe=white,size=fbox,arc=3pt, before upper=\strut, top=-2pt, bottom=-4pt, left=-2pt, right=-2pt, boxrule=0pt}
\newtcbox{\hlsecondarytab}{on line, box align=base, colback=red!10,colframe=white,size=fbox,arc=3pt, before upper=\strut, top=-2pt, bottom=-4pt, left=-2pt, right=-2pt, boxrule=0pt}

\newcommand{\dashifted}{\raisebox{0.5\depth}{\tiny$\downarrow$}}
\newcommand{\uashifted}{\raisebox{0.5\depth}{\tiny$\uparrow$}}
\newcommand{\da}[1]{{\footnotesize\hlsecondarytab{\dashifted{#1}}}}
\newcommand{\ua}[1]{{\footnotesize\hlprimarytab{\uashifted{#1}}}}

\newcommand{\ourmodel}{\textsc{FastFixer}}

\AtBeginDocument{%
  }

\copyrightyear{2024}
\acmYear{2024}
\setcopyright{acmlicensed}\acmConference[ASE '24]{39th IEEE/ACM
International Conference on Automated Software Engineering }{October
27-November 1, 2024}{Sacramento, CA, USA}
\acmBooktitle{39th IEEE/ACM International Conference on Automated Software
Engineering (ASE '24), October 27-November 1, 2024, Sacramento, CA, USA}
\acmDOI{10.1145/3691620.3695062}
\acmISBN{979-8-4007-1248-7/24/10}

\begin{document}
\title{\ourmodel{}: An Efficient and Effective Approach for Repairing Programming Assignments}


\author{Fang Liu$^{1\dag}$, Zhenwei Liu$^{1\dag}$, Qianhui Zhao$^1$, Jing Jiang$^1$$^\ast$, Li Zhang$^1$ \\ Ge Li$^2$, Zian Sun$^1$, Zhongqi Li$^3$, Yuchi Ma$^3$}
\thanks{$^\dag$Fang Liu and Zhenwei Liu contributed equally to this paper. \\ $^{\ast}$Corresponding author.}
\affiliation{%
\institution{$^1$State Key Laboratory of Complex \& Critical Software Environment, School of Computer Science and Engineering, Beihang University, Beijing, China \\ $^2$School of Computer
Science, Peking University, Beijing, China \\
$^3$Huawei Cloud Computing Technologies Co., Ltd, Shenzhen, China}
  \country{}
}

\email{{fangliu, zhaoqianhui, jiangjing, lily, sza}@buaa.edu.cn, meetliuzhenwei@gmail.com} 
\email{lige@pku.edu.cn, {lizhongqi7, mayuchi1}@huawei.com}

\renewcommand{\shortauthors}{Liu et al.}

\begin{abstract}
  Providing personalized and timely feedback for student's programming assignments is useful for programming education. Automated program repair (APR) techniques have been used to fix the bugs in programming assignments, where the Large Language Models (LLMs) based approaches have shown promising results. Given the growing complexity of identifying and fixing bugs in advanced programming assignments, current fine-tuning strategies for APR are inadequate in guiding the LLM to identify bugs and make accurate edits during the generative repair process. Furthermore, the autoregressive decoding approach employed by the LLM could potentially impede the efficiency of the repair, thereby hindering the ability to provide timely feedback. To tackle these challenges, we propose \ourmodel{}, an efficient and effective approach for programming assignment repair. To assist the LLM in accurately identifying and repairing bugs, we first propose a novel repair-oriented fine-tuning strategy, aiming to enhance the LLM's attention towards learning how to generate the necessary patch and its associated context. Furthermore, to speed up the patch generation, we propose an inference acceleration approach that is specifically tailored for the program repair task. The evaluation results demonstrate that \ourmodel{} obtains an overall improvement of 20.46\% in assignment fixing when compared to the state-of-the-art baseline. Considering the repair efficiency, \ourmodel{} achieves a remarkable inference speedup of $16.67\times$ compared to the autoregressive decoding algorithm.  
\end{abstract}



\maketitle

\section{Introduction}
As part of the learning process, it is common for students to make mistakes on their programming assignments. Providing personalized feedback for these mistakes requires a substantial amount of time and effort from teachers and teaching assistants. Together with the rapid growth of computer science, there has been a substantial increase in the demand for programming education \cite{gulwani2018clara, Ahmed2021VerifixVR}, which has given rise to the creation of Massive Open Online Courses (MOOCs) \cite{masters2011moocs}. In such a scenario, the need to provide personalized and timely feedback to a vast number of students has become increasingly urgent. To this end, Automated Program Repair (APR) approaches have received increased attention for providing feedback on student programming assignments \cite{gupta2017deepfix,yasunaga2021break,xia2023chat-repair}.

Early work employed search-based methods \cite{le2011genprog,liu2019tbar} and semantic-based tools to produce the feedback \cite{nguyen2013semfix,mechtaev2016angelix,le2017jfix}. These approaches usually require significant engineering efforts and substantial experience in program analysis. In addition, many of these tools require the integration of customized repair strategies that are specifically designed for the language domain. To overcome these limitations, researchers have introduced learning-based approaches \cite{gupta2017deepfix,pu2016sk_p,yasunaga2021break} that benefit from the advancements in Deep Learning. These approaches leverage an end-to-end methodology to perform repairs, wherein bug-fixing patterns are learned from extensive code repositories. 

In recent years, researchers have started to directly leverage advanced Large Language Models (LLMs) for repairing the student programming assignments \cite{zhang2022repairing,joshi2023repair,zhao2024peer}. These models have shown promising results in student assignment bug fixing, and even outperform previous state-of-the-art APR techniques. 
Most of these approaches focus on the introductory assignment, and it is imperative to develop efficient APR tools for advanced programming assignments \cite{zhao2024peer}. In advanced assignments, the buggy programs are more complex and the errors are more challenging to fix, necessitating a significantly greater amount of effort to assist students in identifying and resolving these errors. To effectively assess the APR tool's performance in the context of programs from advanced programming courses, \citet{zhao2024peer} proposed a dataset named Defects4DS from an advanced programming course, including programs with increased complexity and code lengths. Nearly half of the buggy programs in Defects4DS consist of multiple bugs, with many of them being interrelated, making them more challenging to locate and fix. Along with the dataset, \citet{zhao2024peer} also proposed PaR, a repair framework based on prompting LLM with a multi-source prompt using API access. However, without fine-tuning, it may experience limited performance. Moreover, accessing the LLM API incurs high costs and may result in response latency, impeding the provision of timely feedback in practice.

Given the increased difficulty of locating and repairing bugs in the advanced assignment fixing,
the end-to-end generative LLM fine-tuning approach is considered more practical, and has been used in previous APR work \cite{jiang2023impact,silva2023repairllama}. Specifically, LLMs take the buggy code as input and generate the fixed code from scratch. However, the fine-tuning approach still presents two main challenges. First, when fine-tuning LLMs to fix bugs, they often fail to make necessary edits and instead keep the buggy code unchanged, as seen in Figure \ref{fig:masking_res_example} (``None Masking'' result). 
Effectively guiding the LLM in identifying bug locations and making correct edits during the generative repair process remains a difficult task. Second, LLMs are required to generate all the tokens of the complete corrected program. This generation typically occurs in an autoregressive manner, where tokens are generated one by one. While the inclusion of contextual dependencies among the generated tokens helps ensure the quality of the results, this generation process is slow, especially when the target program is long. For instance, it takes approximately 59.05 seconds to generate a fixed code of 1,779 tokens for CodeLlama-7B \cite{roziere2023codellama} when running on 2 NVIDIA GeForce RTX 3090 GPUs. However, providing quick feedback is essential for enabling effective interaction with students. How to find effective and efficient ways to provide the necessary patches still remains a challenge. These challenges undermine the effectiveness of the approaches, hindering their ability to provide accurate and timely feedback for complex programming assignments.

To address these limitations, we propose \ourmodel{}, a novel approach for repairing programming assignments that incorporates a repair-oriented fine-tuning strategy and an efficient inference algorithm. Specifically, to help the LLM effectively identify and correct the bugs, we propose a novel repair-oriented fine-tuning strategy, aiming to enhance the LLM's attention on learning to generate the patch and its associated context during the fine-tuning process. To generate patches promptly during the inference process, we propose an inference acceleration approach that is specifically tailored for the program repair task. This approach utilizes the buggy code as a draft and employs the LLM to perform parallel verification. The experimental results on the student assignment repair task demonstrate that \ourmodel{} achieves state-of-the-art performance in advanced programming assignment datasets, correctly fixing 312 programs in Defects4DS, obtaining 20.46\% improvement compared to the state-of-the-art approach \cite{zhao2024peer}. 
Considering the repair efficiency, \ourmodel{} is significantly faster than autoregressive decoding, achieving 16.67 $\times$ speed-up. This highlights the effectiveness of our proposed acceleration algorithm.

The contributions of this paper can be summarized as follows:

\begin{itemize}
    \item We design a novel repair-oriented fine-tuning strategy to enhance LLM's attention toward learning how to generate the necessary patches and their relevant contexts. 
    \item We propose an inference acceleration algorithm that is specifically tailored for the program repair task by taking the snippets of the buggy code as a draft to accelerate the inference process. To the best of our knowledge, our research represents the first exploration of inference acceleration in LLM-based APR approaches.
    \item We conducted a comprehensive evaluation of \ourmodel{} and several state-of-the-art APR baselines on repairing buggy programs from advanced programming assignments. 
\end{itemize}

\section{Background and Related Work}

\subsection{Automated Program Repair}
Automated Program Repair (APR) techniques can produce automated patches for the bugs without human efforts, reducing the time and costs associated with debugging. Most early APR works employ search-based or semantic-based methods. Search-based approaches generate patches by searching for correct patches using predefined transformation rules \cite{le2011genprog,liu2019tbar}. 
The semantic-based methods typically repair bugs with symbolic execution \cite{nguyen2013semfix,mechtaev2016angelix}. 
Compared with the general purpose repair, peer solutions or the previous submissions from the same assignment can be utilized for aiding the repair process. \citet{gulwani2018clara} and \citet{pu2016sk_p} generate the patches with clustering and machine learning techniques. \citet{Ahmed2021VerifixVR} align the assignments with the peer reference solution in terms of control flow. 
Sarfgen \cite{wang2018search} generates minimal fixes to repair the incorrect programs by searching for programs in previous student submissions that exhibit similar syntax to the ones containing errors. While these tools can offer strong guarantees for the generated code, they are heavily dependent on their specific domain language. 

In recent years, there has been significant progress in APR using deep learning-based approaches \cite{gupta2017deepfix, jiang2021cure, zhu2021recoder}. Particularly, the LLMs have gained attention and achieved state-of-the-art performance across various APR tasks \cite{xia2023chat-repair,fan2023automated,silva2023repairllama}. 

\subsection{LLM for Automated Program Repair}
The emergence of Large Language Models holds great potential for enhancing program repair \cite{xia2023chat-repair,zhao2024peer,jiang2023impact}. Most LLMs are trained using an autoregressive approach that predicts the next token by considering the context of past tokens within extensive corpora. After training, LLMs are either prompted \cite{zhao2024peer,xia2023chat-repair} or fine-tuned \cite{silva2023repairllama,hossain2024Toggle} to enable their utilization for program repair.

Prompt-based approaches allow for the direct utilization of LLMs without the need for any additional training. 
These tools utilize a predefined prompt, encompassing the task description and the buggy code, to query the LLM and generate a new code snippet. 
\citet{zhang2022repairing} proposed to employ Codex \cite{chen2021evaluatingCodex} for repairing introductory Python programming assignments. \citet{xia2023chat-repair} built an APR system based on ChatGPT \cite{chatgpt} in a conversational style, which utilized instant feedback to generate a patch. Most of the existing APR tools focus on the introductory programming assignments repair. Most recently, \citet{zhao2024peer} proposed an advanced programming assignment dataset Defects4DS, where the programs are more complex and contain several related errors, which are more challenging to locate and fix. 
In this paper, we employ their dataset to evaluate the performance of our model in advanced programming assignment repair.
Along with the dataset, \citet{zhao2024peer} also proposed PaR, a framework for advanced student assignment repair based on LLMs. It incorporates a peer solution selection strategy and a multi-source prompt generation method. 
Without fine-tuning, existing prompt-based tools may encounter limited performance in specific domains.

Fine-tuning approaches update the model parameters by training the model on downstream tasks using corresponding training datasets. 
Researchers have adopted the practice of fine-tuning LLMs for program repair \cite{jiang2023impact,silva2023repairllama,hossain2024Toggle}. For instance, \citet{jiang2023impact} investigated the impact of LLMs on APR, including both the prompt-based and fine-tuning strategies. 
\citet{silva2023repairllama} proposed RepairLLaMa, a method that fine-tunes CodeLlama \cite{roziere2023codellama} with LoRA \cite{lora} for program repair. They also leverage repair-specific code representations by incorporating the repair signals. 
Many of these models rely on identifying the bug location and customizing repair tasks accordingly. However, recognizing the bug location can be challenging in practical scenarios. In \ourmodel{}, we eliminate the need for a separate bug location identification and introduce a novel fine-tuning approach aimed at guiding the model to learn the fixed pattern end-to-end effectively.

\begin{figure*}[t]
  \centering
  \setlength{\abovecaptionskip}{0.1cm}
  \includegraphics[width=0.7\textwidth]{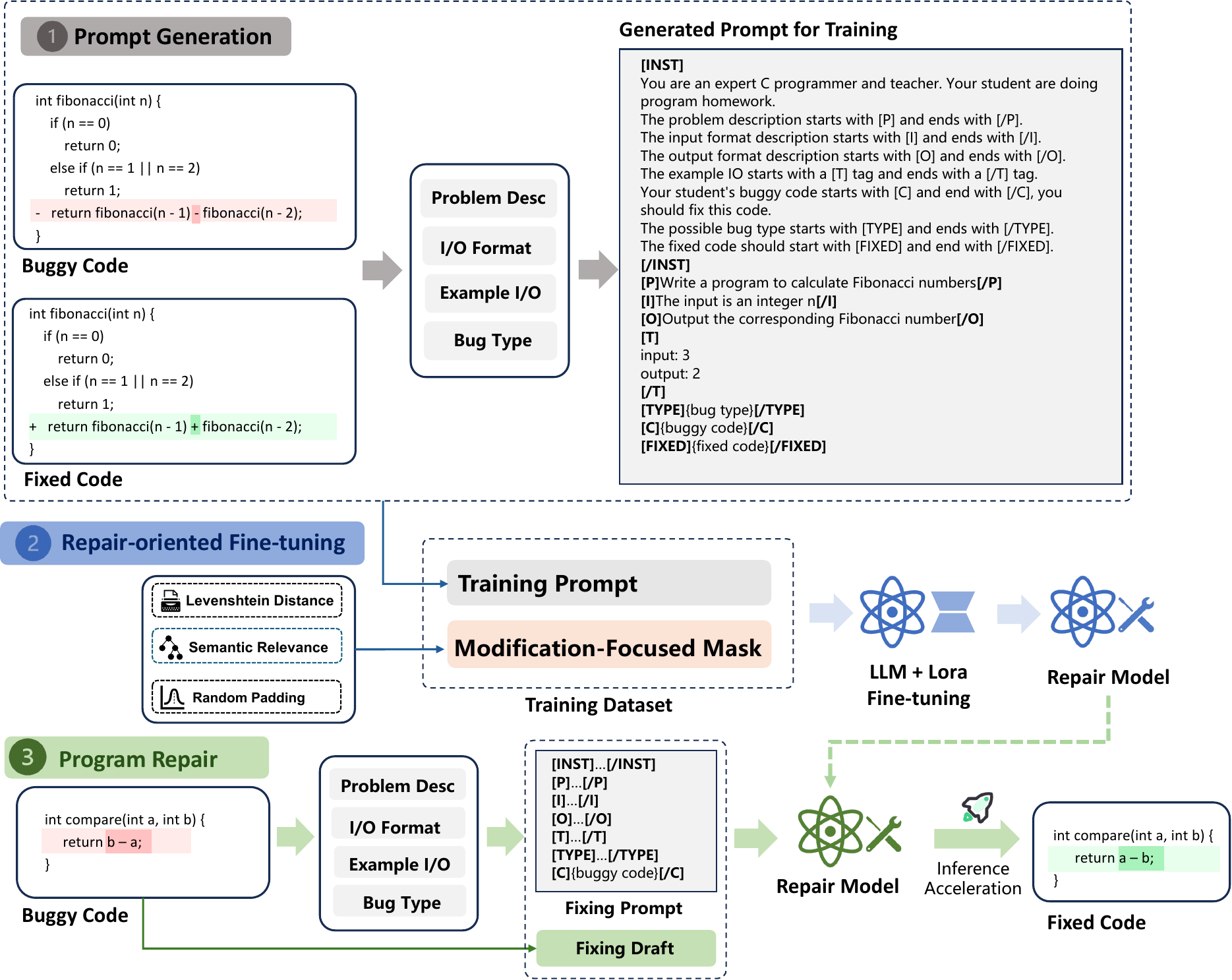}
  \caption{Overview of \ourmodel{}.}
  \label{overview}
  \vspace{-0.3cm}
\end{figure*}

\subsection{Efficient Inference for LLM}\label{Efficient-LLM}
Most existing LLMs employ an AutoregRessive (AR) decoding algorithm, which generates tokens in a step-by-step manner and results in a slower decoding process.
In the era of LLMs, efficient inference becomes a crucial aspect of model deployment and service. To address this challenge, various approaches have been proposed to accelerate the inference, which can mainly be divided into two categories: lossy acceleration and lossless acceleration. 

Lossy acceleration methods focus on training models that exhibit similar behavior and achieve faster inference compared to the target LLM. Some examples of such methods include non-autoregressive decoding \cite{liu2024non,gu2017non}, pruning \cite{frantar2023sparsegpt}, knowledge distillation \cite{sanh2019distilbert}, and quantization \cite{liu2023Llm-qat}. Lossless acceleration approaches aim to speed up the decoding of LLM while preserving its behavior. One extensively studied method involves reducing the number of function calls during decoding. For example, \citet{leviathan2023speculative} proposed the speculative decoding strategy, which employs a smaller model to generate a draft sequence autoregressively with less computational cost, and then utilizes the LLM to verify the draft tokens in parallel. Specifically, LLM will accept contiguous draft tokens that are aligned with its output, discard inconsistent tokens that follow, and generate a new token to fix the first rejected token. 

The speculative decoding framework reduces the overall inference cost of LLM, particularly when a high-quality draft is obtained. Various approaches are proposed to produce a high-quality draft. 
\citet{yang2023LLMA} and \citet{he2023rest} retrieve the draft tokens from the reference document or a datastore. 
\citet{cai2024medusa} introduced multiple heads to predict multiple draft tokens in parallel.
Jacobi decoding approach \cite{santilli2023Jacobi} leverages Jacobi and Gauss-Seidel fixed-point iteration methods to accelerate the inference, which first initializes a draft with padding tokens, and then utilizes the LLM to verify the draft in parallel.
\citet{fu2024lookahead} further improved the Jacobi decoding algorithm by keeping track of the trajectory of Jacobi decoding, generating and caching the n-gram from this trajectory.

Drawing inspiration from these methodologies and considering the unique characteristics of APR tasks, with the buggy code serving as a natural high-quality draft, we propose an inference acceleration method for APR. 

\section{Method}

The overview of our approach is presented in Figure \ref{overview}. 
As mentioned before, the bugs in the advanced programming assignment are challenging to locate by existing fault location techniques. For example, we have attempted to use spectrum-based fault localization (SBFL) to locate bugs, but it fails in most cases because many buggy programs cannot pass any test cases, and the execution coverage of some test cases approaches 100\%. Additionally, multiple (related) errors can exist in one buggy code, further exacerbating the bug location challenge. Therefore, we chose to generate the entire program in an end-to-end manner during the repair, which can work when the bug location is not available.
We divided the procedure into three stages: \textit{Prompt Generation, Repair-oriented Fine-tuning, and Program Repair}.
In \textit{Prompt Generation}, we design a high-quality prompt to provide the model with the necessary context for performing repair. 
In \textit{Repair-oriented Fine-tuning}, we build a training dataset using students' historical submissions and fine-tune the LLM with a modification-focused mask strategy. 
In \textit{Program Repair}, the tuned LLM is used to repair buggy code. In this stage, we design an inference acceleration strategy using the buggy code as a draft to accelerate the repair process.

\subsection{\textbf{Prompt Generation}}

Prompts play a crucial role in LLMs, acting as a key bridge connecting external task requirements with the model's internal knowledge.
In the domain of APR, prompts should provide essential context for the repair procedure, guiding the model in identifying errors within the code while providing hints for generating accurate and personalized feedback.
To achieve this, we meticulously design a prompt leveraging the abundant signals within the educational domain to guide the LLM in repairing bugs within the student assignment. In particular, we utilize the fact that programming assignments typically provide \textit{detailed Problem Descriptions (Problem Desc)}, \textit{specifications of Input and Output Format (I/O Format)}, and \textit{Example IOs} (which are not present in the test cases) to assist students in comprehending the problem. 
Additionally, based on preliminary experiments, we discovered that including the bug type in the prompt enhances LLM's ability to generate more precise patches during fine-tuning. We further incorporate this information into the prompt by designing heuristic rules to identify and collect several common and easily recognizable \textit{Bug Type} from the buggy code, including compile error, timeout error, and presentation error. 
Specifically, we identify compile errors and gather this information from compiler outputs, assess the occurrence of timeout errors by analyzing program execution time, and detect presentation errors by comparing the format of outputs with the expected one.

\subsection{Repair-oriented Fine-tuning}

Through end-to-end repair fine-tuning, the LLM struggles to effectively identify and correct errors, often simply reproducing the original buggy code without making the necessary modifications.
This undermines the effectiveness of the model's repair capability. 
To mitigate this issue, we propose a novel repair-oriented fine-tuning approach with a modification-focused masking strategy to guide the LLM to learn how to generate the necessary patch and its associated context.

\subsubsection{\textbf{Problem Definition}}
The goal of APR is to learn the transformation between the buggy code \(X = p_{1}, p_{2}, \ldots, p_{m}\) and the fixed code \(Y = q_{1}, q_{2}, \ldots, q_{n}\), where \(p_{i}\) and \(q_{i}\) represent the \(i\)-th statement of the buggy and fixed codes, respectively. Training an APR model is to fit a mapping function, which can be defined as:
\(f\left(p_{1}, p_{2}, \ldots, p_{m}, q_{1}, q_{2}, \ldots, q_{i - 1} \right) = q_{i}\), where the model takes the buggy code and the partially fixed code generated so far as input, producing the following fixed code \(q_{i}\). However, the correct and buggy codes typically have minor differences, where \(p_{i} = q_{i}\) in most cases, which forms a pattern of preserving the existing, albeit incorrect, code.
Consequently, the model could ``replicate'' the original buggy code instead of correcting it if we directly fine-tune the model with the original buggy-fix code pairs. To this end, we propose a modification-focused masking strategy to guide the model to focus on learning the code snippets that require essential modifications during the fine-tuning process.

\begin{figure}[t]
  \centering
  \setlength{\abovecaptionskip}{0.1cm}
  \includegraphics[width=0.5\textwidth]{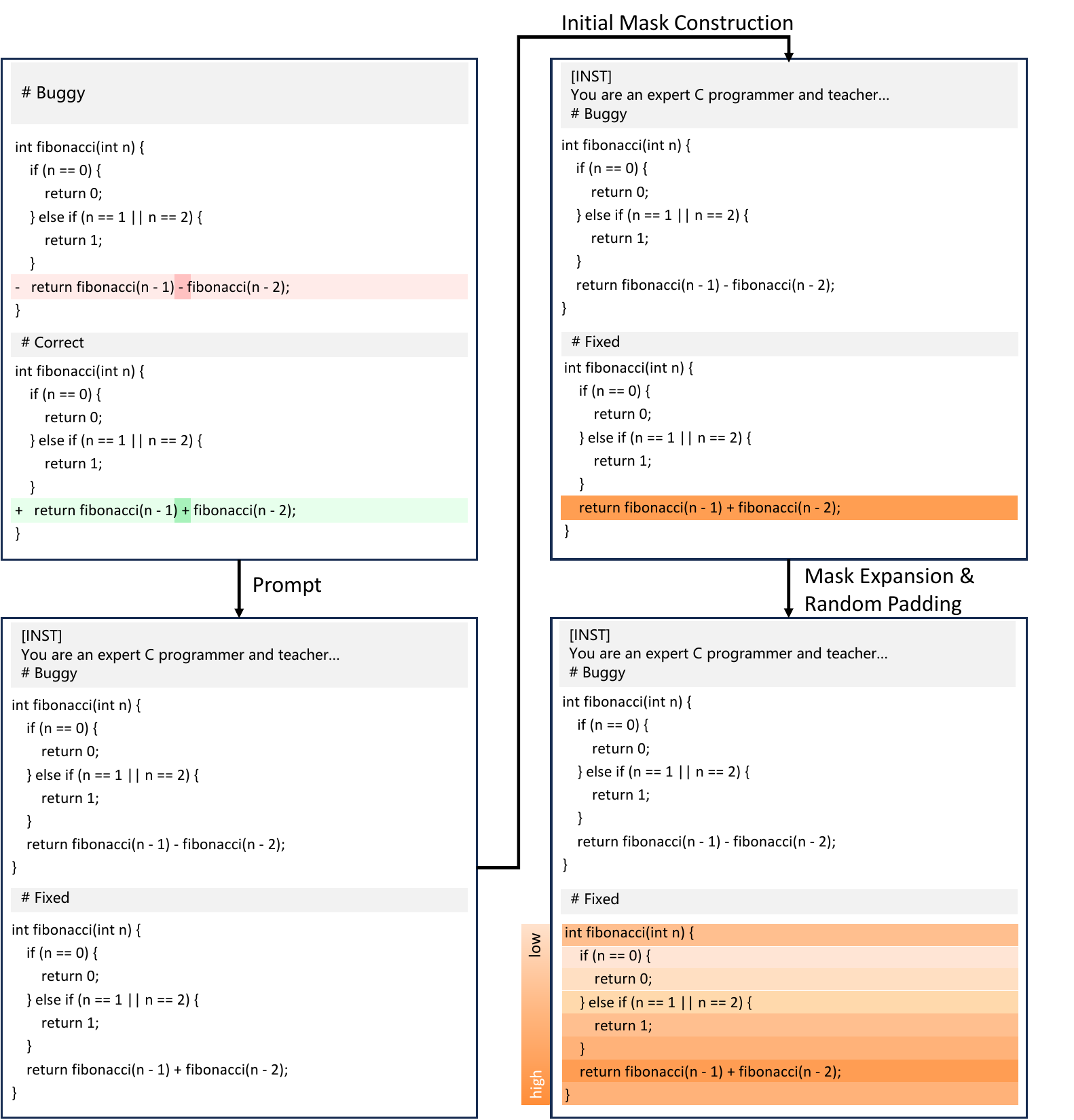}
  \caption{Modification-focused mask construction process.} \label{fig:mask}
  \Description{The Process of Building Modification-Focused Mask.}
  \vspace{-0.4cm}
\end{figure}

\subsubsection{\textbf{Modification-Focused Mask Construction}}
To enhance the efficiency of the training procedure by placing more emphasis on identifying and processing code segments that require repair, we design a method to create a special mask that highlights the code snippets in the fixed code closely associated with the patch, which we call the modification-focused mask.
Figure \ref{fig:mask} illustrates the process of constructing the Modification-Focused Mask, which involves the following three steps.

\noindent \textbf{(1) Initial Mask Construction.} We first compare the buggy code \(X\) and the fixed code \(Y\) to identify the code difference statements \(Y_m\) from \(Y\) based on the Levenshtein distance \cite{levenshtein1966binary}. For example, in Figure \ref{fig:mask}, we can identify the statement \lstinline|return fibonacci(n - 1)+fibonacci(n - 2);| as \(Y_m\). The weight for each statement in \(Y_m\) is set to 1.
Meanwhile, we also identify the set of variables \(V_m\) in \(Y_m\), such as \verb|n|, and the set of function calls \(F_m\) in \(Y_m\), such as \verb|fibonacci| for the next step.

\noindent \textbf{(2) Mask Expansion.} Except for the directly modified statements \(Y_m\), it is crucial to pay attention to statements that are semantically related to \(Y_m\) during the repair tuning process. Therefore, we extend the initial mask to include the following statements \(Y_e\): (1) statements that assign values to variables in \(V_m\), (2) statements that define functions in \(F_m\), and (3) statements that reside within the same control flow block as the statement in \(Y_m\). The weight for each statement in \(Y_e\) is determined by the relevance score \(\text{sim}(e, m)\), which is calculated based on its distance with the corresponding modification parts: 
\begin{equation}\label{eq:dist-formula}
\text{sim}(e, m) = \frac{1}{1 + \log(\text{dist}(e, m)) + 1}
\end{equation}
where \(e \in Y_e, m \in Y_m\), and \(dist\) is the Levenshtein distance between the two statements. The weight of each statement \(e\) in \(Y_e\) is calculated as the following formula:
\begin{equation}\label{eq:dist-formula}
\text{weight}(e) = \max(1,\sum\limits_{m\;in\;Y_m}(\text{sim}(e,m)))
\end{equation}

\noindent \textbf{(3) Random Padding.} To prevent the model from being overly aggressive in its modification operations, we finally assign a random weight below a predefined threshold \(\sigma\) to the rest code statements.

Following the above steps, we construct a Modification-Focused Mask vector \(K = k_{1}, k_{2}, \ldots, k_{n}\), where \(k_i \in (0, 1)\), and \(\sum_{i=1}^{n} k_i=1\). \(k_i\) represents the attention that needs to be paid when generating \(q_i\), which is produced by normalizing the modification weights obtained in the preceding three steps.

\vspace{-0.1cm}

\subsubsection{\textbf{Repair-oriented Fine-Tuning}}
Based on the above msking approach, we generate the Modification-Focused Mask vector \(K\) for all the training instances. Then we use them to bias the fine-tuning loss, aiming to enhance the repair ability of the LLM.
Specifically, we incorporate the Modification-Focused Mask vector as the weight when computing the Repair-oriented Fine-Tuning loss function \(\mathcal{L}_{RoFT}\) as follows:
\begin{equation}\label{eq:masked-loss}
\mathcal{L}_{RoFT} = \sum_{i=1}^{n} \mathcal{L}_{origin}(q_i|X, q_1, q_2, ..., q_{i-1}) \cdot k_i
\end{equation}
where \(\mathcal{L}_{origin}\) is the original cross-entropy loss, and $\mathcal{L}_{RoFT}$ is computed by the weighted sum of the original loss, where the weight \(k_i\) for each target statement \(q_i\) is obtained from our proposed Modification-Focused Mask vector.

\vspace{-0.1cm}

\subsection{Program Repair with Inference Acceleration}
When utilized for programming assignment repair, the well-tuned LLM has to generate all the tokens for the complete fixed code as the bug location is unknown. If we use autoregressive inference, in which the LLM decodes one new token in each inference step, there will be a significant inference cost, especially when dealing with longer programs.
To tackle this challenge, we introduce an inference acceleration approach tailored specifically for the APR task.

\vspace{-0.1cm}

\subsubsection{\textbf{Problem Definition}}
During the LLM's inference process for bug fixing, the objective is to generate the fixed code $Y=(y_1, y_2, ..., y_N)$, token by token, based on the buggy code $X=(x_1, x_2, ..., x_M)$:
\begin{equation}
    y_{i} = \arg\max p_{\theta}\left(y_{i}\;|\;X,y_{1}:y_{i-1}\right)
\end{equation}
where \(x_i\) is the \(i\)-th token vector in \(X\), \(y_i\) is the \(i\)-th token vector in \(Y\), \(M\) and \(N\) are the number of tokens in \(X\) and \(Y\), respectively.
In order to obtain the complete fixed code $Y$, it is necessary to perform $N$ inference steps:

\begin{equation}
    \left\{
    \begin{aligned}
        y_{1} &= \arg\max p_{\theta}\left(q_{1}\;|\;X\right) \\
        y_{2} &= \arg\max p_{\theta}\left(q_{2}\;|\;X,y_{1}\right) \\
        &\vdots \\
        y_{N} &= \arg\max p_{\theta}\left(q_{n}\;|\;X,y_{1}:y_{N-1}\right)
    \end{aligned}
    \right.
\end{equation}

In fact, the LLM makes predictions for each input position at every inference step:
\begin{equation}\label{eq:actual-decoding}
  p_{\theta}\left(x_{2},..., x_{M},y_{1},..., y_{i}\;|\;x_1, x_2, ..., x_M,y_{1},y_2, ..., y_{i-1}\right)
\end{equation}
This means that, while making prediction for \(y_i\), LLM also makes prediction for $x_1, ..., x_M, y_1:y_{i-1}$, where $y_{1}:y_{i-1}$ has been predicted in previous decoding steps. Although this indicates that the LLM is capable of parallel computation, the cumulative dependency among tokens requires the use of autoregressive decoding, thereby preventing true parallelization.
Thus, the key to acceleration lies in breaking the constraints imposed by the cumulative dependency among the target tokens. 
In the program repair task, we propose to create a high-quality draft based on the buggy code, and then use LLM to verify and revise the draft.

\begin{algorithm}[t]
  \caption{Inference Acceleration with Buggy Code as a Draft}
  \label{alg:fix-acceleration}
  \begin{algorithmic}[1]
  \setstretch{1.1}
  \STATE \textbf{Input:} buggy code $X$, InputPrompt $I$, $N_{max}$, $f$
  \STATE \textbf{Output:} fixed code $Y$
  \STATE \textbf{Initailize:} 
  $stop \gets False$
  $end \gets 1$
  $\hat{Y} \gets \emptyset$
  $o \gets N_{max}$
  $\boldsymbol{x} \gets I$
  \WHILE {$stop = False$}
    \STATE $y_{end}^{'}:y_{o}^{'} \gets \text{DraftGeneration}(X,\hat{Y})$
    \STATE $y_{end}:y_{o} \gets \arg\max p_{\theta}(y_{end}:y_{o}\;|\;\boldsymbol{x}, y_{end}^{'}:y_{o}^{'})$
    \STATE $k \gets \text{LongestMatchingPrefix}(y_{end}:y_{o}, y_{end}^{'}:y_{o}^{'})$
    \STATE $\boldsymbol{x} \gets \boldsymbol{x} + y_{end}:y_{end+k+1}$
    \STATE $\hat{Y} \gets \hat{Y} + y_{end}:y_{end+k-1}$
    \STATE $end \gets end + k$
    \STATE $stop \gets \text{IsEnd}(\hat{Y})$
    \IF {$stop = True$}
      \STATE $Y = \hat{Y}$
      \RETURN $Y$
    \ENDIF
    \STATE $stop, \hat{Y}, end, \boldsymbol{x} \gets ErrorHandling(\hat{Y}, end, \boldsymbol{x}, f)$
  \ENDWHILE
  \STATE $Y = \hat{Y}$
  \RETURN $Y $
  \end{algorithmic}
\end{algorithm}

\subsubsection{\textbf{Parallel Decoding in APR}}
Since there are often only minor differences between the buggy code \(X\) and the fixed code \(Y\), we propose the inference acceleration algorithm for APR by using fragments of the buggy code as a draft to accelerate the repair process.
Specifically, we first construct a high-quality draft $y_{1}^{'}:y_{o}^{'}$ based on the buggy code, and then perform one autoregressive inference step with the LLM to get the output $y_{1}:y_{o}$:
\begin{equation}
  y_{1}:y_{o} = \arg\max p_{\theta}\left(y_{1}:y_{o}\;|\;\textbf{x},y_{1}^{'}:y_{o}^{'}\right)
\end{equation}
By comparing $y_{1}:y_{o}$ with $y_{1}^{'}:y_{o}^{'}$, the longest matching prefix $y_{1}:y_{k}$ is extracted as the accepted output as it matches the LLM's autoregressive decoding outputs. The unmatched tokens following this prefix are discarded and then we use the LLM to generate the following token(s) of current statement. This verification process ensures that the decoding algorithm produces output identical to the autoregressive decoding process. 
By doing this, we can get $k$ tokens in parallel, which accelerates the decoding process. 
If $k = o$, the decoding process is fully parallel; conversely, if $k = 1$, the algorithm degenerates into a completely serial decoding. 
The detailed workflow is shown in Algorithm \ref{alg:fix-acceleration}, which operates by executing the following steps iteratively:

\noindent \textbf{(1) Draft Construction.} Given the buggy code \(X\) and previously accepted draft tokens $\hat{Y}$, we employ \textit{DraftGeneration} function to generate the draft tokens $y_{end}^{'}:y_{o}^{'}$ by first aligning the previously accepted code $\hat{Y}$ and $X$, and then concatenating $\hat{Y}$ with the rest part of $X$. Initially, \(\hat{Y}=\emptyset\), the draft tokens are initialized with $X$. 

\noindent \textbf{(2) Draft Verification and Acceptance.} Then LLM is used to generate the verified output candidates $y_{end}:y_{o}$ with a single forward pass. Then we apply \textit{LongestMatchingPrefix} function to compute the longest matching prefix $k$ between $y_{end}:y_{o}$ and $y_{end}^{'}:y_{o}^{'}$. The matched $k$ tokens will be accepted as part of the final output, and will be appended to both $\hat{Y}$ and LLM's input \(x\).
Then we employ \textit{IsEnd} function to determine when to stop the algorithm. Specifically, if the [EOS] token, \textit{i.e.}, End Of Sequence symbol, is included in $\hat{Y}$ or the maximum length limit $N_{max}$ is reached, the algorithm stops and returns the verified code up to the [EOS] token or the maximum length. However, if not all the draft tokens are accepted, the algorithm proceeds to the next procedure.

\noindent \textbf{(3) Error Handling.} In this stage, our algorithm uses the LLM to generate the following \(f\) tokens with standard autoregressive decoding based on the buggy code \(X\) and the previously verified tokens $\hat{Y}$, and append the generated tokens to$\hat{Y}$. This ensures the output is identical to the original AR decoding process.
Then the algorithm moves to the new decoding process.

\begin{table*}[t]
  \centering
  \setlength{\abovecaptionskip}{0.1cm}
  \caption{Dataset statistics of Defects4DS-L. CGC is complex grammatical components, CF is custom function, and M-Array stands for multi-dimensional array \cite{zhao2024peer}. For the buggy type, SE is the semantic error, TLE is the time limit error, CE is the compile error, and PE is the presentation error. }
  \resizebox{\textwidth}{!} {
  \begin{tabular}{lccccccccccccc}
  \toprule
   & \multirow{2}{*}{\textbf{\# Buggy Files}} & \multicolumn{2}{c}{\textbf{\# Lines}} & \multicolumn{2}{c}{\textbf{\# of Tokens}} & \multicolumn{3}{c}{\textbf{\# CFG}} & \multicolumn{4}{c}{\textbf{\# Bug Type}} & \multirow{2}{*}{\textbf{\# Fixed Files}} \\ \cmidrule(r){3-4} \cmidrule(r){5-6} \cmidrule(r){7-9} \cmidrule(r){10-13}
   & & \textbf{Avg.} & \textbf{Med.} & \textbf{Avg.} & \textbf{Med.} & \textbf{M-Array} & \textbf{Pointer} & \textbf{Struct} & \textbf{SE} & \textbf{TLE} & \textbf{CE} & \textbf{PE} & \\ 
   \midrule
   Train & 40,663 & 98.05 & 82 & 965.45 & 797 & 6,630 & 18,101 & 7,408 & 34,928 & 1,817 & 2,082 & 1,836 & 13,231 \\
   Test & 4,623 & 92.95 & 78 & 903.15 & 760 & 731 & 2,082 & 893 & 3,790 & 110 & 120 & 493 & - \\ 
  \bottomrule
  \label{tab:dataset}
  \end{tabular}
  }
  \vspace{-0.4cm}
\end{table*}

\vspace{-0.2cm}

\section{Evaluation}

\subsection{Research Questions}

In this work, we aim to answer the following research questions:

\noindent \textbf{RQ1: Overall Repair Performance.}
  This research question aims to evaluate the overall performance of \ourmodel{} in repairing bugs within the programming assignments when compared to the state-of-the-art APR tools. We further analyze \ourmodel{}'s performance in repairing different bug types.
  
\noindent \textbf{RQ2: Impact of Masking Strategy.}
  This question evaluates the influence of various masking strategies on the effectiveness of \ourmodel{} employed in the fine-tuning process. 
  
\noindent \textbf{RQ3: Repair Efficiency.}
  This research question examines the inference efficiency of our proposed acceleration strategy by quantifying the speed-up it provides when compared to the autoregressive decoding approach. Moreover, we also compare our acceleration method with a widely used inference acceleration approach \cite{fu2024lookahead}.

\subsection{Datasets}

Defects4DS \cite{zhao2024peer} contains 682 submissions writing in C language from 4 programming assignments of a Data Structure course in a college, which offers an opportunity to enhance the evaluation of advanced program assignment repair models.
\citet{zhao2024peer} evaluated both their model (PaR) and several state-of-the-art APR tools on the Defects4DS dataset. In our experiments, we also leveraged Defects4DS as a benchmark to evaluate the performance of \ourmodel{} in RQ1.

To fine-tune our model, it is essential to construct a well-suited fine-tuning dataset that not only aligns with the evaluation task but also encompasses a sufficient number of programs. We notice that the authors of \citet{zhao2024peer} have released a supplementary dataset for the assignment, collected from the Data Structure course. This dataset includes students' submissions for 18 programming assignments spanning over three years.
In the evaluation, to make a comparison with SOTA baselines, we first evaluate \ourmodel{} on the Defects4DS proposed in \citet{zhao2024peer} in RQ1, where the submissions are from only one year. We curate the fine-tuning dataset by extracting the submissions from the other two years and use it to fine-tune \ourmodel{} and evaluate it on Defects4DS. 
However, Defects4DS is limited to submissions from only 4 assignments and exclusively preserves programs with semantic errors. To conduct a systematic analysis of performance across different problem difficulties and bug types, we expanded our experiment to evaluate the performance on a larger test set. Specifically, we reorganized the full dataset, where the test set encompasses all 18 assignments from one year, while the submissions from the other two years serve as the fine-tuning dataset. We name this dataset as Defects4DS-L. The fine-tuning data instances were constructed by pairing each incorrect submission with a correct submission from the same student on the same problem. Subsequently, we filtered out pairs where the edit distance exceeds 10, as we consider such cases to represent restructuring rather than repair behaviors.

Defects4DS-L allows us to cover a wide range of task types with the test programs, including bugs of various types, not limited to semantic errors alone. Each assignment of Defects4DS and Defects4DS-L includes test cases provided by the instructor to assess the correctness of students' submissions. 
The detailed dataset information is shown in Table \ref{tab:dataset}, including the number of buggy files and fixed files, the buggy program size, and the distribution of grammatical components and bug types. 
It is important to note that the program size and the number of complex grammatical components are significantly higher when compared to the introductory assignment dataset \cite{itsp}, indicating a greater level of complexity and difficulty in fixing the bugs in Defects4DS-L.

\vspace{-0.3cm}

\subsection{Metrics}

\subsubsection{Repair Capability}
We use the following metrics to measure the model's repair capability:
\begin{itemize}[leftmargin=*]
    \item \textbf{Number of Fixed Programs} is the number of fixed programs for each problem, which can pass all the test cases within the problem.
    \item \textbf{Fix Rate} is calculated by dividing the number of fixed programs by the total number of buggy programs for each problem.
\end{itemize}

\subsubsection{Inference Efficiency}
We measure the inference efficiency with the following metrics, all of which are computed on two NVIDIA GeForce RTX 3090 GPUs. These metrics are calculated throughout the decoding process of a single target program with batch size 1, and are averaged over the entire test set.
\begin{itemize}[leftmargin=*]
    \item \textbf{Speedup} is the average ratio of the time taken to repair one buggy code before and after applying the inference acceleration algorithm ($time_{before}/time_{after}$) over the whole test set.
    \item \textbf{Tokens/s} represents the average number of tokens generated per second during the inference. 
    \item \textbf{Step Efficiency} measures the average reduction in inference steps by computing the average ratio of inference steps required to repair a buggy code before and after applying the acceleration ($step_{before}/step_{after}$) over the whole test set.
    \item \textbf{Avg. Time} is the average time (second) the LLM requires to repair a buggy code over the whole test set.
\end{itemize}

\subsection{\textbf{Compared Baselines}}
\subsubsection{Repair Performance}
For the repair ability evaluation, we compare \ourmodel{} with several state-of-the-art APR approaches, including two prompt-based LLM baselines (ChatRepair \cite{xia2023chat-repair} and PaR \cite{zhao2024peer}), and a fine-tuning based LLM baseline (RepairLlama \cite{silva2023repairllama}).
\begin{itemize}[leftmargin=*]
    \item \textbf{ChatRepair} \cite{xia2023chat-repair} is a conversation-driven bug-fixing approach that utilizes test failure information as feedback to iteratively repair bugs with ChatGPT. Since it has not been open-sourced, we replicate it based on the approach described in the paper. 
    \item \textbf{PaR} \cite{zhao2024peer} utilizes a peer solution strategy to retrieve the closely related peer program, and combines it with multiple sources of information to create the prompt for repair. There are two variations of PaR with different backbones, namely PaR-ChatGPT and PaR-CodeLlama. In RQ1, we use their proposed dataset, Defects4DS, to evaluate the performance of \ourmodel{} and baselines, where the corresponding results of PaR are copied from their paper. 
    \item \textbf{RepairLlama} \cite{silva2023repairllama} fine-tunes CodeLlama \cite{roziere2023codellama} with repair-specific code representations by incorporating the repair signals. Given that the precise buggy signals, such as bug location, are not available in our fine-tuning dataset, we reproduce RepairLlama by fine-tuning CodeLlama on our dataset with their IR1 $\times$ OR1 setting. In this setting, the input is the buggy function whereas the output is the fixed function.    
\end{itemize}

\subsubsection{Inference Acceleration}
For inference acceleration, we compare our strategy with the following decoding approaches:
\begin{itemize}[leftmargin=*]
    \item \textbf{Autoregressive decoding} (AR) is the typical decoding approach employed by existing LLMs, which generates tokens one by one.
    
    \item \textbf{Lookahead decoding} \cite{fu2024lookahead} accelerates inference by generating n-grams from the Jacobi iteration trajectory, and using LLM to verify.  
    We replicated their methodology based on their public artifact\footnote{https://github.com/hao-ai-lab/LookaheadDecoding}, and integrated it into our framework with the best parameter setting.\footnote{Multiple parameter configurations were evaluated, and the optimal combination was identified as \(LEVEL = 5\), \(W = 7\), and \(G = 7\).}
    \item \textbf{RepairLlama-IR4\(\times\)OR4} \cite{silva2023repairllama} generates diffs instead of full code. We assume the presence of an oracle error localizer and use RepairLlama-IR4\(\times\)OR4 as our baseline to facilitate a comparison of repair efficiency.
\end{itemize}

\subsection{\textbf{Implementation Details}}

We select CodeLlama-7B \cite{roziere2023codellama} as our backbone LLM following \citet{silva2023repairllama}, and fine-tune it using LoRA \cite{lora} based on the Pytorch framework and the Transformer package. 
The pre-trained model was downloaded from the Huggingface platform.\footnote{https://huggingface.co/codellama} 
For Modification-Focused Mask construction, we use CParser library\footnote{https://pypi.org/project/cparser/} for code parsing. we tested various thresholds $\sigma$ from 0.1 to 0.8 and ultimately selected the best setting of 0.6 as our threshold. 

During the fine-tuning stage, we employed CodeLlama's default maximum input length of 4,096 and set the maximum decoding length, $N_{max}$, as 4,192 based on the statistical analysis of the dataset. We employed the Adam optimizer with a learning rate of 1e-4 and a batch size of 4. Furthermore, We experimented with different values for the Lora rank—specifically 4, 8, and 16, and eventually selected 8 as the optimal rank.
The acceleration factor $f$ in Error Handling was set to 5 based on preliminary testing.
Both the model and the dataset were processed using the fp16. The model is trained with 8,000 steps on a 64-core server equipped with two NVIDIA GeForce RTX 3090 GPUs and 64GB of DDR5 RAM. Evaluations were also performed on this server configuration.

\section{Results} 

\begin{table*}[t]
  \centering
  \setlength{\abovecaptionskip}{0.1cm}
  \caption{Comparison of \ourmodel{} with baselines on bug fixing capability on Defects4DS. The numbers in parentheses following the problem ID represent the number of submissions.}
  \begin{tabular}{llllllll} 
  \toprule
  \multirow{2}{*}{\textbf{Model}} & \multirow{2}{*}{\textbf{Backbone}} & \multicolumn{5}{c}{\textbf{\# of Fixed Programs}} & \multirow{2}{*}{\textbf{Avg. LED}} \\
  \cmidrule{3-7}
   &  & \textbf{Prob.1 (123)} & \textbf{Prob.2 (108)} & \textbf{Prob.3 (197)} & \textbf{Prob.4 (254)} & \textbf{Overall (682)} \\
  \midrule
  RepairLlama & CodeLlama (7B) & \textbf{91} & 14 & 44 & 93 & 242 & 1.05 \\
  ChatRepair & ChatGPT & 84 & 12 & 6 & 86 & 188 & 16.58 \\
  PaR-CodeLlama & CodeLlama (34B) & 74 & \underline{33} & \underline{53} & 65 & 225& 33.84 \\
  PaR-ChatGPT & ChatGPT & \textbf{91} & 25 & 29 & \underline{114} & \underline{259} & 24.61\\
  \ourmodel{} & CodeLlama (7B) & \underline{90} \da{1.10\%} & \textbf{37} \ua{12.12\%} & \textbf{59} \ua{11.32\%} & \textbf{126} \ua{10.53\%} & \textbf{312} \ua{20.46\%} & 3.74 \\
  \bottomrule
  \end{tabular}
  \vspace{-0.3cm}
  \label{tab:performance_comparison}
\end{table*}

\subsection{\textbf{RQ1: Overall Repair Performance}}
\subsubsection{Comparison with SOTA baselines}
To answer this RQ, we conducted a comprehensive comparison of \ourmodel{} against four state-of-the-art techniques: ChatRepair \cite{xia2023chat-repair}, RepairLlama \cite{silva2023repairllama}, and two variations of PaR \cite{zhao2024peer} (PaR-ChatGPT and PaR-CodeLlama) on Defects4DS dataset. Table \ref{tab:performance_comparison} presents the results on four assignments within Defects4DS. As seen from the results, PaR-ChatGPT performs the best among all the baselines, and \ourmodel{} outperforms all the baselines by a large margin, achieving an overall improvement of 20.46\% when compared to PaR-ChatGPT. PaR-ChatGPT retrieves the most semantically similar peer solution as a reference and incorporates it into the prompt. This enables the model to directly analyze the inherent relationship between the reference solution and the buggy code, facilitating the repair process. However, the effectiveness of their model relies on the availability of high-quality peer solutions. In contrast, \ourmodel{} trains the open-sourced LLM to learn the repair patterns using historical submissions collected from other years. As a result, it achieves better robustness, does not depend on peer solutions, and is not limited by the accessibility of the model API.
Comparing \ourmodel{} with RepairLlama, the main difference lies in the fine-tuning strategy. Our model is fine-tuned using a novel repair-oriented approach, which focuses on learning the code snippets that necessitate crucial modifications. The performance improvement further demonstrates the effectiveness of our modification-focused masking and fine-tuning strategy. 

Regarding the performance in repairing different assignments, \ourmodel{} consistently achieves the best performance across Prob.2 to Prob.4, while also achieving comparable performance with the best baseline in Prob.1. Most of the bugs that appear in Prob.1 are about output formatting errors, which are relatively easy to fix \cite{zhao2024peer}. Therefore, most of these APR techniques demonstrate notable performance in resolving them.
Notably, RepairLlama and PaR-CodeLlama, which both utilize the CodeLlama base model, excel on Prob.3, surpassing the methods based on ChatGPT, \textit{i.e.}, PaR-ChatGPT and ChatRepair. However, their performance is less competitive in Prob.4 when compared to PaR-ChatGPT. 
Meanwhile, \ourmodel{} consistently delivers superior results for all of these problems, with notable improvements exceeding 10\% on Prob.2-Prob.4. This demonstrates the robustness and effectiveness of our model. 

The last column presents the average \textit{Line-level Edit Distance (LED)} between the fixed code and the buggy code. Compared to fine-tuning based approaches (RepairLlama and FastFixer), prompt-based baselines tend to make more modifications, which might increase the potential for introducing ambiguities that could hinder students' learning process. 

\subsubsection{Performance in repairing different types of bugs}
We further analyze the performance of \ourmodel{} in-depth in repairing different types of bugs on Defects4DS-L, and make a comparison with RepairLlama. The results are shown in Table \ref{tab:bug_types}. Among the submissions, semantic errors are the most frequent and are typically more challenging to locate and repair \cite{zhao2024peer}. For these semantic errors, \ourmodel{} can achieve an improvement of 15.99\% compared to RepairLlama. For the remaining three error types, the presentation errors are easier to fix, most of which only involve formatting issues and do not involve deep-level code logic. Interestingly, the number of fixed programs by our model and RepairLlama is identical, with 299 being successfully fixed by both models. For the compile and timeout errors, \ourmodel{} consistently outperforms the baseline. These results demonstrate that by encouraging the model to focus on the modified code snippets and their relevant context during fine-tuning, \ourmodel{} can better capture the code semantics and repair logic, ultimately enhancing its repair capability. This underscores the effectiveness of our fine-tuning strategy.

\begin{table}[h]
  \centering
  \setlength{\abovecaptionskip}{0.1cm}
  \caption{Performance (\# of Fixed Programs) in repairing different types of bugs on Defects4DS-L.}
  \resizebox{1.0\linewidth}{!}{
  \begin{tabular}{lll} 
  \toprule
  \textbf{Bug Types (\# of submissions)} & \textbf{RepairLlama} & \textbf{\ourmodel{}} \\
   \midrule
   Compile Error (120) & 27 & \textbf{32} \ua{18.52\%} \\
   Timeout Error (110) & 35 & \textbf{43}\ua{22.86\%} \\
   Presentation Error (493) & \textbf{351} & \textbf{351} \\
   \midrule
   Semantic Error (3,900) & 1113 & \textbf{1291} \ua{15.99\%} \\
  \bottomrule
  \end{tabular}}
  \label{tab:bug_types}
  \vspace{-0.3cm}
\end{table}

\subsection{\textbf{RQ2: Impact of Masking Strategy}}
To evaluate the effectiveness of the proposed Modification-Focused Masking strategy in fine-tuning, we conduct a comprehensive experimental study on the Defects4DS-L to investigate how different masking strategies impact the performance of code repair. In addition to our adopted masking strategy (M4), we also introduce three other masking approaches (M1-M3) for comparison:

\begin{itemize}[leftmargin=*]

  \item \textbf{Modification Only (M1):} 
  This masking approach only computes loss for the modified statements, while ignoring the remaining part. Specifically, the loss weights for the modified statements are set to 1, and others are 0.
  The core idea of this masking strategy is to guide the model focusing exclusively on modifications.
  
  \item \textbf{Modification + Random Padding (M2):} 
  Building on the M1, this strategy further introduces random weights to the unmodified parts of the code. 
  This strategy aims to find a balance between preserving the original code and enabling effective repairs. This helps prevent the model from becoming excessively aggressive in its modifications.
  
  \item \textbf{Modification + Semantic Relatedness (M3):} 
  In contrast to assigning random weights to other statements, this strategy assigns weights based on the semantic relatedness of the code. The goal is to guide the model in learning not only the modified code segments but also their semantically related counterparts, allowing the model to acquire a more comprehensive understanding of the repair context, thereby improving the correctness of repair.
  
  \item \textbf{Modification + Semantic Relatedness + Random Padding (M4):} 
  This strategy combines the benefits of M3 and M2 by introducing random padding on the foundation of M3. This not only enhances the model's semantic learning capability but also prevents it from making overly aggressive modifications, which is adopted in our framework.

\end{itemize}

Table \ref{tab:masking_strategies} illustrates the results of these four masking strategies (M1, M2, M3, M4) on bug repair across 18 problems. The difficulty level for each problem is determined by considering both the nature of the task and the proficiency required for its successful implementation. To make a comparison, we also include the results of the original fine-tuning strategy without any masking intervention in the ``Origin'' column. To measure the extent of modifications, we present the average \textit{Line-level Edit Distance (LED)} between the fixed code and the buggy code for each masking strategy in the last row.

As seen from the results, M3 and M4 consistently achieve superior performance across a wide range of difficulty levels and for the majority of the problems analyzed. Specifically, M4 performs the best and shows an improvement of 4.13\% in the fix rate compared to the original settings.
Strategy M1, which solely focuses on the modified code segments, exhibits worse performance across the majority of problems, and is unable to fix any bugs in Prob.6. This suggests that solely concentrating on modified segments without considering any additional context may result in suboptimal repair performance. The large edit distance of M1 further indicates its excessively aggressive modifications, introducing numerous unnecessary alterations that negatively impact repair performance.
By applying random weights to the contextual tokens, M2 can yield improvements compared to M1.
Moreover, M3 and M4 demonstrate improvements over M2 for most problems.
This observation underscores the beneficial impact of incorporating semantic relatedness into the repair process. 
By integrating semantic relatedness, these strategies enable models to gain a deeper understanding of the patch's context. This, in turn, allows for a more comprehensive acquisition of information relevant to the bug fixing, resulting in the generation of more precise patches.

\begin{table}[t]
\begin{center}
    \setlength{\abovecaptionskip}{0.1cm}
    \caption{Fixed rate of different masking strategies. Avg. LED denotes the average Line-level Edit Distance between the fixed code and the buggy code.}
  \footnotesize
  \begin{tabular}{llccccc}
    \toprule
    \multirow{2}{*}{{\textbf{Prob ID}}} & \multirow{2}{*}{\textbf{Difficulty}} &  \multicolumn{5}{c}{\textbf{Fix Rate}} \\
    \cmidrule{3-7}
     & & \textbf{Origin} & \textbf{M1} & \textbf{M2} & \textbf{M3} & \textbf{M4} \\
     \midrule
    Prob.1 & easy & 70.26\% & 45.45\% & 67.69\% & \textbf{70.77\%} & 67.18\% \\
    Prob.2 & medium & 19.08\% & 15.65\% & 21.97\% & 21.97\% & \textbf{25.43\%} \\
    Prob.3 & easy & 25.21\% & 22.16\% & \textbf{31.51\%} & 28.57\% & 30.67\% \\
    Prob.4 & medium & 38.54\% & 21.25\% & 40.13\% & \textbf{43.63\%} & 37.81\% \\
    Prob.5 & medium & 19.38\% & 13.76\% & 25.19\% & 25.19\% & \textbf{42.68\%} \\
    Prob.6 & easy & 33.71\% & 0.00\% & \textbf{38.72\%} & 36.22\% & 27.91\% \\
    Prob.7 & medium & 35.78\% & 28.33\% & 32.11\% & 35.32\% & \textbf{38.07\%} \\
    Prob.8 & hard & 27.02\% & 9.61\% & 31.99\% & 32.92\% & \textbf{34.78\%} \\
    Prob.9 & medium & 48.48\% & 60.92\% & 50.76\% & \textbf{56.06\%} & 55.30\% \\
    Prob.10 & hard & 39.42\% & 3.06\% & \textbf{43.51\%} & 38.94\% & 39.66\% \\
    Prob.11 & medium & 53.77\% & 48.87\% & 58.22\% & 60.27\% & \textbf{61.64\%} \\
    Prob.12 & hard & 28.53\% & 0.00\% & 26.65\% & \textbf{30.72\%} & 26.65\% \\
    Prob.13 & hard & 24.46\% & 0.00\% & 28.80\% & \textbf{29.89\%} & 28.26\% \\
    Prob.14 & hard & 18.57\% & 1.87\% & 21.43\% & \textbf{30.00\%} & 22.14\% \\
    Prob.15 & hard & 18.18\% & 2.53\% & \textbf{30.58\%} & 22.31\% & 24.79\% \\
    Prob.16 & hard & 41.30\% & 1.30\% & 46.38\% & \textbf{45.65\%} & 39.13\% \\
    Prob.17 & medium & 27.55\% & 12.40\% & 35.60\% & \textbf{38.08\%} & 37.77\% \\  
    Prob.18 & hard & 24.19\% & 16.18\% & \textbf{29.68\%} & 25.44\% & 27.43\% \\
    \textbf{Overall} & - &33.01\% & 12.59\% & 36.77\% & 37.08\% & \textbf{37.14\%} \\
    \midrule
    \textbf{Avg. LED} & - & 6.07 & 24.11 & 7.64 & 8.26 & 7.80 \\ 
    \bottomrule
  \end{tabular}
  \label{tab:masking_strategies}
  \end{center}
  \vspace{-0.3cm}
\end{table}

Figure \ref{fig:over-modify} provides an example of the repair results of different masking strategies. 
As depicted, the only necessary correction involves changing \verb|return top==99| to \verb|return top>=99|. Fine-tuning models using a None Masking strategy fails to capture and correct this error. Conversely, models fine-tuned with the M1 strategy are capable of identifying and fixing this specific error. 
However, the sole focus on the modified section results in overly aggressive modifications, introducing numerous unnecessary alterations.
While the M4 strategy can correctly repair this error with minimal modifications, without introducing unnecessary changes. Consequently, we have selected M4 as our preferred strategy for implementation.

Subsequently, We conducted a manual inspection of 50 randomly-sampled repairs that were successfully repaired by both M3 and M4. The results reveal that M3 leads to 6 (12\%) instances of excessively aggressive modification, while only 2 (4\%) instances were found in M4. This suggests that M4 tends to repair the code with minor extraneous changes, reducing the potential for introducing ambiguities that could hinder students' learning. Therefore, we selected M4 as preferred strategy.

\begin{figure}
  \centering
  \setlength{\abovecaptionskip}{0.1cm}
  \includegraphics[width=0.5\textwidth]{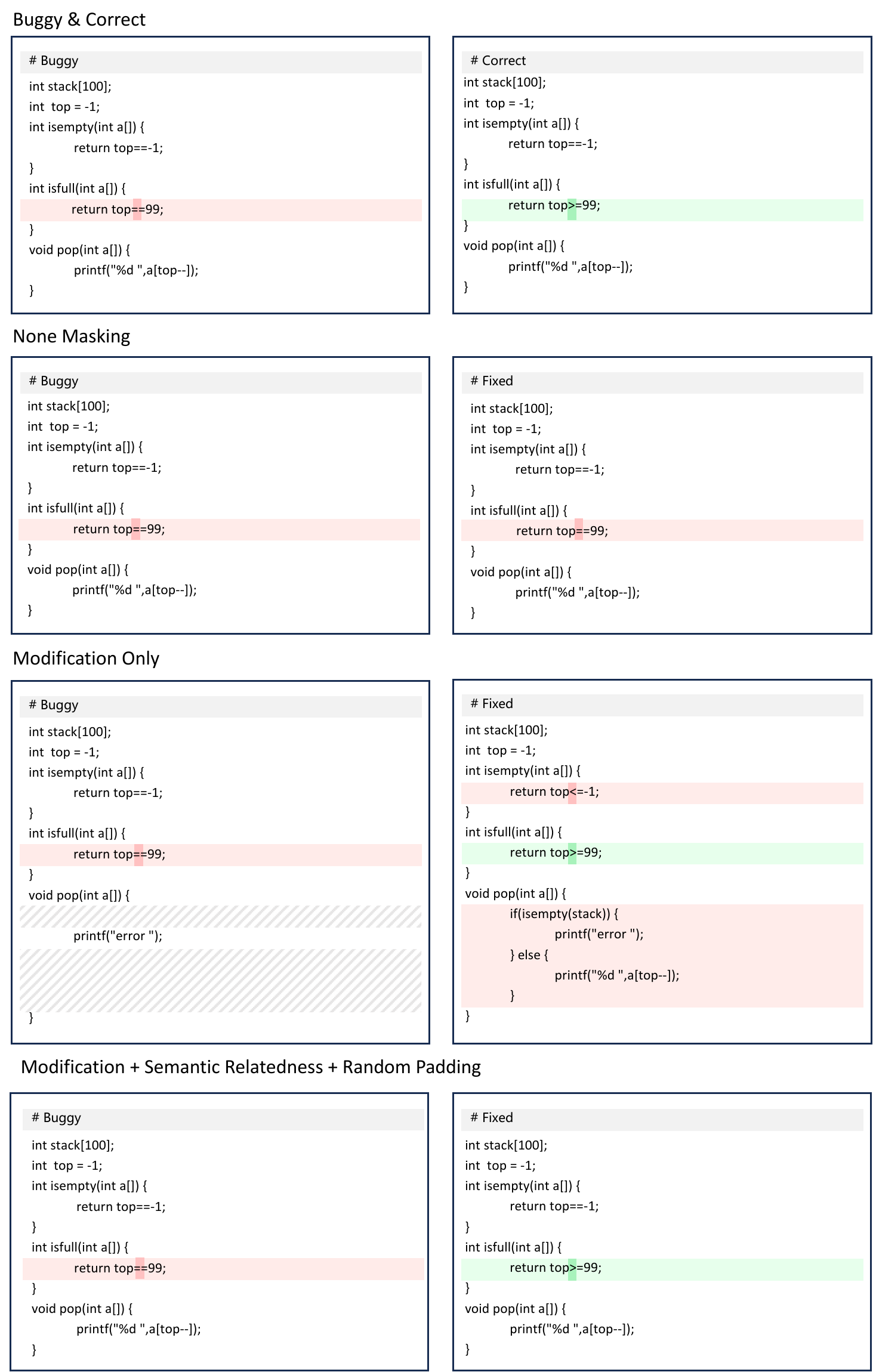}
  \caption{Bug fixing examples of various masking strategies.} \label{fig:over-modify}
  \Description{An Example of Over-modification in the Repair Process.}
  \label{fig:masking_res_example}
  \vspace{-0.3cm}
\end{figure}

\subsection{\textbf{RQ3: Repair Efficiency}}
\subsubsection{Comparison with baselines}

Table \ref{tab:infer_results} presents the inference efficiency metrics of \ourmodel{} and the baselines, \textit{i.e.}, autoregressive (AR), Lookahead decoding strategies, and RepairLlama (IR4\(\times\)OR4). 
As seen from the results, the average time required to repair a buggy code is less than 6 seconds, much smaller than baseline decoding algorithms. Moreover, the average number of tokens generated per second is significantly greater than that of the baselines, achieving an improvement of over $10\times$. 
In terms of \textit{Step Efficiency}, \ourmodel{} achieves an average reduction of $147\times$ in the number of inference steps required to repair a program compared to AR. 
As mentioned in Section \ref{Efficient-LLM}, the quality of the draft is a crucial factor in accelerating inference. Our draft is derived from the buggy code, which shares a high similarity with the target program. As a result, only minimal efforts are required to verify and revise the draft, significantly reducing the inference steps required to repair a program and ultimately reducing the overall inference time. In the case of longer programs with minor bugs, the reduction can even exceed $1000\times$. 
In contrast, the draft of Lookahead decoding or other acceleration algorithms like \cite{santilli2023Jacobi,leviathan2023speculative} is either randomly initialized or generated by models. Consequently, increased efforts are necessary when there is a significant discrepancy between the draft and the target output, leading to more inference steps and higher inference time costs.

Regarding the \textit{Speedup}, \ourmodel{} achieves $16.67\times$ speedup compared to the AR decoding approach, where the widely used Lookahead decoding approach achieves less than $2\times$. The improvement in speedup is smaller compared to step efficiency. This could be attributed to the Key-Value (KV) cache missing, which will be elaborated on in the next subsection.
Moreover, the speedup achieved by \ourmodel{} surpasses that of RepairLlama-IR4\(\times\)OR4, despite the latter assuming the availability of error-localization and only generating modifications. The reason is that even only generating diffs, there still remains token-level overlap between the fixed code and the buggy code, especially when the model fails to fix bugs and outputs the original buggy code. Since RepairLlama-IR4\(\times\)OR4 generates modifications token-by-token autoregressively, \ourmodel{} requires fewer average inference steps compared to RepairLlama-IR4\(\times\)OR4.

\begin{table}[h]
  \centering
  \setlength{\abovecaptionskip}{0.1cm}
  \caption{Inference efficiency comparison.}
  \resizebox{1.0\linewidth}{!}{
  \begin{tabular}{lcccc}
  \toprule
  \textbf{Method} & \textbf{Speedup} & \textbf{Tokens/s} & \textbf{Step Efficiency} & \textbf{Avg. Time (s)} \\ 
  \midrule
  AR       & 1\(\times\)    & 33.64  & 1\(\times\)      & 26.67  \\ 
  Lookahead   & 1.80\(\times\) & 37.79  & 2.24\(\times\)   & 21.85  \\ 
  RepairLlama-IR4\(\times\)OR4 & 6.69\(\times\) & 35.64	& 15.1\(\times\) & 11.69 \\
  \midrule
  \ourmodel{}   & \textbf{16.67\(\times\)} & \textbf{423.6} & \textbf{147.03\(\times\)} & \textbf{5.91}  \\ 
  \bottomrule
  \end{tabular}}
  \label{tab:infer_results}
  \vspace{-0.3cm}
\end{table}

\begin{figure}[h]
  \centering
  \setlength{\abovecaptionskip}{0.1cm}
  \begin{minipage}{0.45\textwidth}
      \centering
      \includegraphics[width=\textwidth]{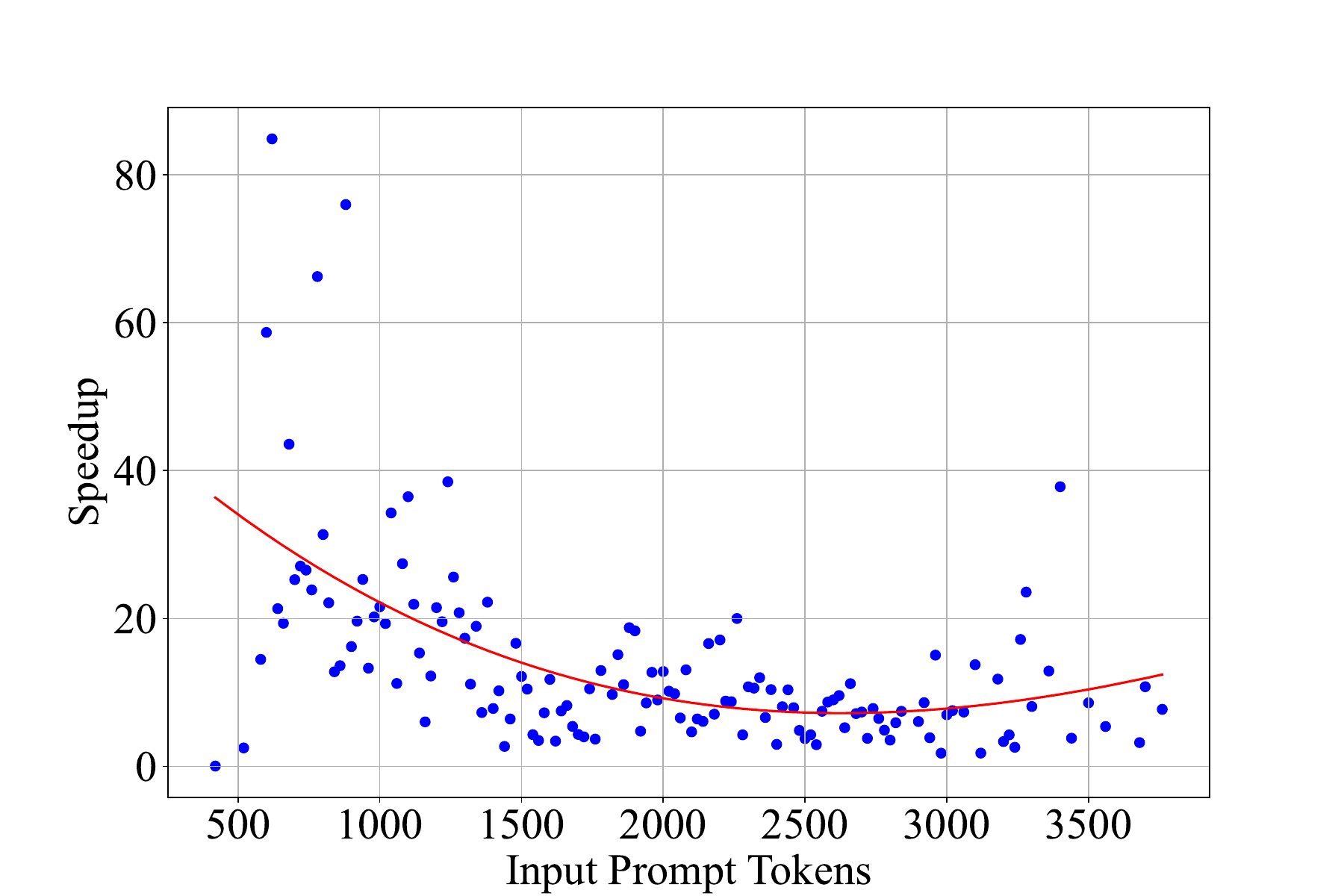}  
      \caption{Inference speedup of \ourmodel{} with different buggy code length.}
      \label{fig:speed}
  \end{minipage}\hfill
  \begin{minipage}{0.45\textwidth}
      \centering
      \includegraphics[width=\textwidth]{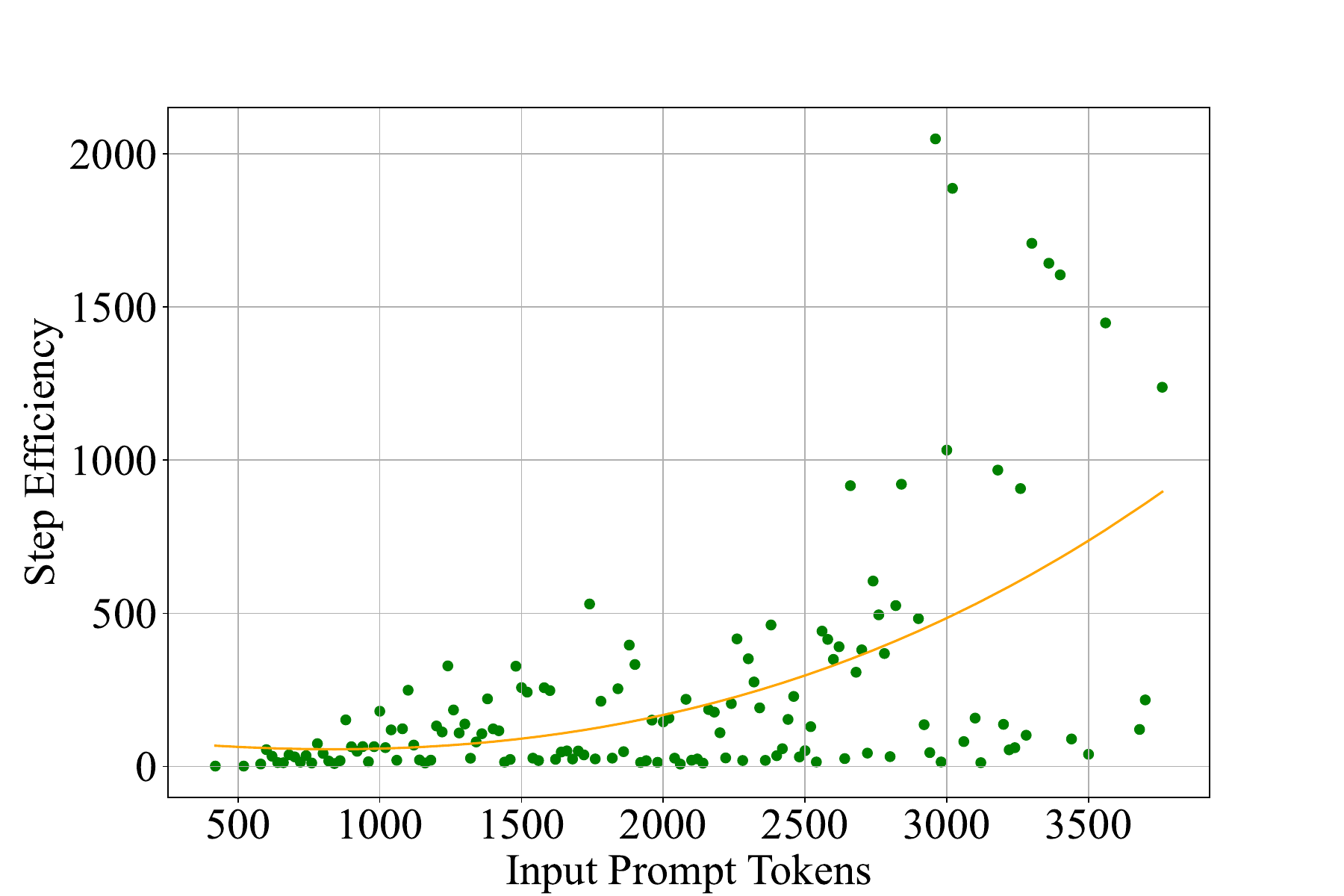}  
      \caption{Inference step efficiency of \ourmodel{} with different buggy code length.}
      \label{fig:step}
  \end{minipage}
  \vspace{-0.3cm}
\end{figure}

\subsubsection{Impact of code length}
To further investigate the impact of the buggy code length on the inference efficiency, we analyze the generation speedup and step efficiency of \ourmodel{} with different buggy code lengths. In Figure \ref{fig:speed} and Figure \ref{fig:step}, we illustrate the speedup and step efficiency of \ourmodel{} on the test set of Defects4DS-L in relation to the increasing buggy code length, with the curves being fitted with a quadratic function.
We observe that step efficiency increases linearly as the length of the buggy code increases. This improvement can be attributed to our accelerated inference using a high-quality draft and parallel verification, resulting in nearly constant-level decoding steps. In contrast, the inference steps of AR decoding increase linearly with the increase in code length. Consequently, the inference speedup achieved by \ourmodel{} increases as the length of the buggy code extends.

Notably, the generation speedup initially decreases and then rises when input prompt tokens exceed 2500. This fluctuation is due to two primary factors: step efficiency and the time cost per step. 
In each verification iteration, a new draft is generated and added to the end of the input prompt, which may differ from the previous version. This could result in a KV cache missing, requiring the LLM to recalculate the key and value vector, leading to additional time costs compared to the AR decoding.
In the beginning, the primary factor leading to a decrease in speedup as the code length increases is the additional time cost caused by the KV-cache missing. However, when the length of the code grows beyond 2500 tokens, the substantial enhancement in step efficiency becomes the predominant factor, thereby increasing the generation speedup.

\begin{figure}
  \centering
  \setlength{\abovecaptionskip}{0.1cm}
  \includegraphics[width=0.45\textwidth]{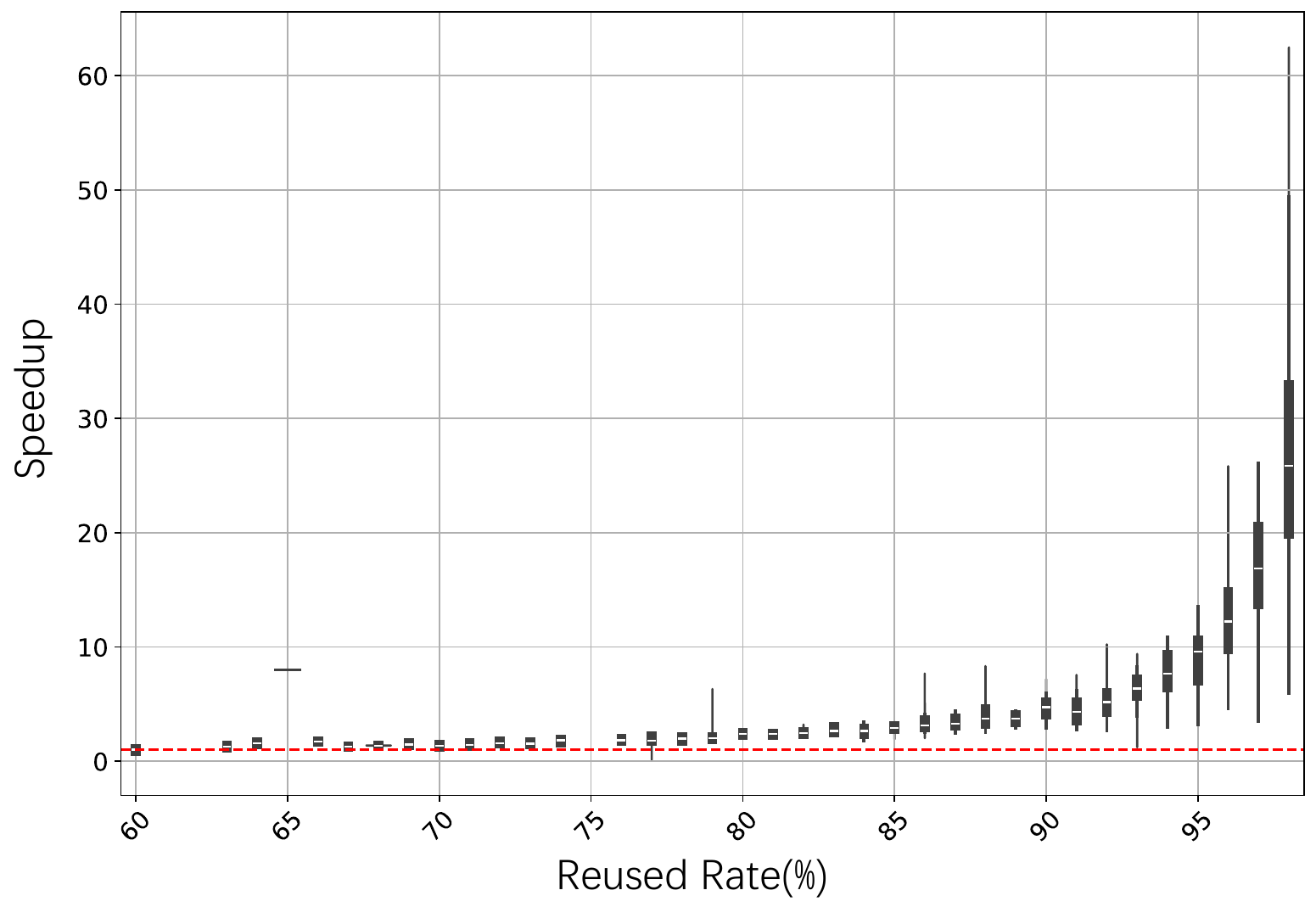}
  \caption{Speedup with different reused code rate.} \label{fig:over-modify}
  \label{fig:speedup_vs_reused_rate}
  \vspace{-0.3cm}
\end{figure}

\subsubsection{Impact of reused code rate}
Given that our decoding acceleration algorithm works under the assumption that fixed code and buggy code are similar. We further analyze the impact of the reused code rate on the speedup. We present a graph of speedup vs \% of re-used code in Figure \ref{fig:speedup_vs_reused_rate}. As seen in the figure, the speedup increases as the code reuse rate rises. Most repairs can reuse more than 60\% buggy code, and there are instances where all the buggy code is reused. This typically occurs when new code snippets need to be added or when the repair fails, resulting in the output of the original buggy code. The high-speedup outliers are primarily the result of timeout errors in the original buggy code, where significant speedup occurs even with a low reused rate.

\section{Threats to Validity}
\textbf{Threats to internal validity} relate to the potential data leakage issue. This could happen when the students' submissions are included in the training data of our backbone model, Code Llama \cite{roziere2023codellama}. However, as \citet{zhao2024peer} mentioned, students are required to complete the assignments independently and the submissions in the test data are not uploaded to public repositories. Furthermore, the submissions utilized for fine-tuning and testing in our evaluation were collected from different years, ensuring that the students and the associated buggy code are distinct. As a result, we believe that the risk of data leakage is minimal. Another internal threat relates to the strong assumption of the similarity between the buggy code and the fixed code. To ensure that buggy code can be repaired with a manageable number of edits rather than requiring a near-complete rewrite, and to provide students with targeted guidance, the discrepancy between the correct and buggy code in Defects4DS(-L) is limited to <=10 lines. However, this assumption may not hold in some cases, particularly when students make incorrect assumptions or fail to write code in a modular way. In such instances, the resulting speedup may not be as substantial.

\noindent\textbf{Threats to external validity} relate to the generalization capability of \ourmodel{}. In RQ1, we assess our model using the Defects4DS, an advanced student assignment benchmark proposed by \citet{zhao2024peer}.
We further expand the test set to incorporate additional programs (Defects4DS-L), enabling a comprehensive evaluation across various assignment difficulties and bug types. Thus, we consider the associated threat to be minimal. 
In this paper, we primarily focus on bug-fixing within the education domain, where historical submissions are typically available. However, our fine-tuning strategy and acceleration algorithm are versatile and can be extended to general bug-fixing scenarios following fine-tuning. In such cases, the fine-tuning dataset can be obtained from open-source repositories with similar tasks/bug-types. For instance, RepairLlama \cite{silva2023repairllama} utilizes Megadiff \cite{monperrus2021megadiff} as their fine-tuning dataset and is tested on Defects4J \cite{just2014defects4j} and HumanEval-Java \cite{jiang2023impact}. While the performance improvement may not be as significant as the education domain when the relevance between the fine-tuning and test datasets is lower, it could still achieve considerable performance as demonstrated by RepairLlama.

\noindent \textbf{Threats to construct validity} relate to the metrics used in our evaluations. For bug fixing, we assess the repairs by executing the program on the test suite provided by the instructors. Although evaluating program correctness through tests may not be as robust as formal verification, the use of tests for correctness is widely used in both APR and educational domains. For inference acceleration, we adopted several metrics widely used in related work to evaluate the inference time and steps.

\section{Conclusion}
We have introduced \ourmodel{}, an effective and efficient framework for providing feedback on advanced programming assignments. It utilizes a repair-oriented fine-tuning approach by guiding the LLM to learn how to generate the necessary patch and its associated context, thus enhancing its repair capability. We also introduce a parallel decoding algorithm tailored for the APR task to accelerate the inference process. Experimental results on thousands of real student submissions demonstrate that \ourmodel{} is effective and outperforms existing APR tools in terms of correctness and efficiency.
To facilitate further research, the dataset and code scripts are available at \url{https://github.com/LiuFang816/FastFixer}.

\begin{acks}
This work is supported by the National Science and Technology Major Project Grant No. 2021ZD0112900, the National Natural Science Foundation of China Grants Nos. 62302021, 62177003, and 62332001, CCF-Huawei Populus Grove Fund CCF-HuaweiSE202302, and the Self-determined Research Funds of State Key Laboratory of Complex \& Critical Software Environment SKLSDE-2023ZX-15. 
\end{acks}

\bibliographystyle{ACM-Reference-Format}
\bibliography{sample-base}

\end{document}